\documentclass{article}
\usepackage{graphicx}
\usepackage{amsmath}

\input{tcilatex}

\begin{document}

{\huge \ }

{\huge New Strings for Old Veneziano }

\ \ \ \ \ \ \ \ \ \ \ \ \ \ \ \ \ \ \ \ \ \ \ \ {\huge Amplitudes :}

\ \ \ \ \ \ {\huge I Analytical Treatment}

$\ \ \ \ \ \ \ \ \ \ \ \ \ \ \ \ \ \ \ \ \ \ \ \ \ \ \ \ \ \ \ \ \ $

\ \ \ \ \ \ \ \ \ \ \ \ \ \ \ \ \ \ \ \ \ \ \ \ \ \ \ A.L. Kholodenko

\textit{375 H.L.Hunter Laboratories, Clemson University, Clemson, }

\textit{SC} 29634-0973, USA

\bigskip \textbf{Abstract}

The bosonic string theory evolved as an attempt to find a physical/quantum
mechanical model \ capable of reproducing Euler's beta function (Veneziano
amplitude) and its multidimensional analogue. The multidimensional analogue
of beta function was studied mathematically for some time from different
angles by mathematicians such as Selberg,Weil and Deligne among many others.
The results of their studies apparently were not taken into account in
physics literature on string theory. In a recent publication [IJMPA 19
(2004) 1655] an attempt was made to restore the missing links. The results
of \ this publication are incomplete, however, since no attempts were made
at reproduction of known spectra of both open an closed bosonic strings or
at restoration of the underlying model(s) reproducing such spectra.
Nevertheless, discussed in this publication the existing mathematical
interpretation of the multidimensional analogue of Euler's beta function as
one of the periods associated with the corresponding differential form
''living'' on the Fermat-type (hyper) surfaces, happens to be crucial for
restoration of the quantum/statistical mechanical model reproducing such
generalized beta function. Unlike the traditional formulations, this model
is supersymmetric. Details leading to restoration of this model will be
presented in the forthcoming Parts 2-4 of our work. They are devoted
respectively to the group-theoretic, symplectic and combinatorial treatments
of this model. In this paper the discussion is restricted mainly to study of
analytical properties of the multiparticle Veneziano and Veneziano-like
(tachyon-free) amplitudes. In the last case, we demonstrate that the
Veneziano-like amplitudes alone (with parameters adjusted accordingly) are
capable of reproducing known spectra of both open and closed bosonic
strings. The choice of parameters is subject to some constraints dictated by
the mathematical interpretation of these amplitudes as periods of
Fermat-type (hyper)surfaces considered as complex manifolds of Hodge type.

\bigskip


\textit{Subj Class}.: String theory, scattering amplitudes


multidimensional beta and hypergeometric functions, Fermat hypersurfaces,

period integrals for Hodge-type manifolds

\bigskip

\bigskip

\pagebreak

\ \ \ 

\section{\protect\bigskip Brief review of the Veneziano amplitudes}

\bigskip In 1968 Veneziano [1] postulated 4-particle scattering amplitude $%
A(s,t,u)$ given (up to a common constant factor) by 
\begin{equation}
A(s,t,u)=V(s,t)+V(s,u)+V(t,u),  \tag{1.1}
\end{equation}
where 
\begin{equation}
V(s,t)=\int\limits_{0}^{1}x^{-\alpha (s)-1}(1-x)^{-\alpha (t)-1}dx\equiv
B(-\alpha (s),-\alpha (t))  \tag{1.2}
\end{equation}
is Euler's beta function and $\alpha (x)$ is the Regge trajectory usually
written as $\alpha (x)=\alpha (0)+\alpha ^{\prime }x$ with $\alpha (0)$ and $%
\alpha ^{\prime }$ being the Regge slope and the intercept, respectively. In
the case of space-time metric with signature $\{-,+,+,+\}$ the Mandelstam
variables $s$, $t$ and $u$ entering the Regge trajectory are defined by [2] 
\begin{eqnarray}
s &=&-(p_{1}+p_{2})^{2},  \TCItag{1.3} \\
t &=&-(p_{2}+p_{3})^{2},  \notag \\
u &=&-(p_{3}+p_{1})^{2}.  \notag
\end{eqnarray}
The 4-momenta $p_{i}$\ are constrained by the energy-momentum conservation
law leading to relation between the Mandelstam variables: 
\begin{equation}
s+t+u=\sum\limits_{i=1}^{4}m_{i}^{2}.  \tag{1.4}
\end{equation}
Veneziano [1] noticed\footnote{%
To get our Eq.(1.5) from Eq.7 of Veneziano paper, it is sufficient to notice
that his $1-\alpha (s)$ corresponds to ours -$\alpha (s).$} that to fit
experimental data the Regge trajectories should obey the constraint 
\begin{equation}
\alpha (s)+\alpha (t)+\alpha (u)=-1  \tag{1.5}
\end{equation}
consistent with Eq.(1.4) in view of definition of $\alpha (s).$

\textbf{Remark 1.1}.The Veneziano condition, Eq.(1.5),\ can be rewritten in
a more general form. Indeed, let $m,n,l$ be some integers such that $\alpha
(s)m+\alpha (t)n+\alpha (u)l=0$, then by adding this equation to Eq.(1.5) we
obtain, $\alpha (s)\tilde{m}+\alpha (t)\tilde{n}+\alpha (u)\tilde{l}=-1,$or,
more generally, $\alpha (s)\tilde{m}+\alpha (t)\tilde{n}+\alpha (u)\tilde{l}+%
\tilde{k}\cdot 1=0.$ Both equations have been studied extensively in the
book by Stanley [3] and play major role in developments to be presented in
this \ work and in Parts 2-4 which follow.

Veneziano noticed that with help of the constraint, Eq.(1.5), the amplitude $%
A(s,t,u)$ can be equivalently rewritten as follows 
\begin{equation}
A(s,t,u)=\Gamma (-\alpha (s))\Gamma (-\alpha (t))\Gamma (-\alpha (u))[\sin
\pi (-\alpha (s))+\sin \pi (-\alpha (t))+\sin \pi (-\alpha (u))].  \tag{1.6}
\end{equation}
The Veneziano amplitude looks strikingly similar to that suggested a bit
later by Virasoro [4]. The latter (up to a constant) is given by 
\begin{equation}
\bar{A}(s,t,u)=\frac{\Gamma (a)\Gamma (b)\Gamma (c)}{\Gamma (a+b)\Gamma
(b+c)\Gamma (c+a)}  \tag{1.7}
\end{equation}
with parameters $a=-\frac{1}{2}\alpha (s),$etc$.$ also subjected to the
constraint: 
\begin{equation}
\frac{1}{2}\left( \alpha (s)+\alpha (t)+\alpha (u)\right) =-1.  \tag{1.8}
\end{equation}
Use of the formulas 
\begin{equation}
\Gamma (x)\Gamma (1-x)=\frac{\pi }{\sin \pi x}  \tag{1.9a}
\end{equation}
and 
\begin{equation}
4\sin x\sin y\sin z=\sin (x+y-z)+\sin (y+z-x)+\sin (z+x-y)-\sin (x+y+z) 
\tag{1.9b}
\end{equation}
permits us to rewrite Eq.(1.7) in the alternative form (up to unimportant
constant): 
\begin{eqnarray}
\bar{A}(s,t,u) &=&[\Gamma (-\frac{1}{2}\alpha (s))\Gamma (-\frac{1}{2}\alpha
(t))\Gamma (-\frac{1}{2}\alpha (u))]^{2}\times  \notag \\
&&[\sin \pi (-\frac{1}{2}\alpha (s))+\sin \pi (-\frac{1}{2}\alpha (t))+\sin
\pi (-\frac{1}{2}\alpha (u))].  \TCItag{1.10}
\end{eqnarray}
Although these two amplitudes look deceptively similar, mathematically, they
are markedly different. Indeed, by using Eq.(1.6) conveniently rewritten as 
\begin{equation}
A(a,b,c)=\Gamma (a)\Gamma (b)\Gamma (c)[\sin \pi a+\sin \pi b+\sin \pi c] 
\tag{1.11}
\end{equation}
and exploiting the identity 
\begin{equation*}
\cos \frac{\pi z}{2}=\frac{\pi ^{z}}{2^{1-z}}\frac{1}{\Gamma (z)}\frac{\zeta
(1-z)}{\zeta (z)}
\end{equation*}
after some trigonometric calculations the following result is obtained: 
\begin{equation}
A(a,b,c)=\frac{\zeta (1-a)}{\zeta (a)}\frac{\zeta (1-b)}{\zeta (b)}\frac{%
\zeta (1-c)}{\zeta (c)},  \tag{1.12}
\end{equation}
provided that 
\begin{equation}
a+b+c=1.  \tag{1.13}
\end{equation}
For the Virasoro amplitude, apparently, no result like Eq.(1.12) can be
obtained. As the rest of this paper demonstrates, the differences between
the Veneziano and the Virasoro amplitudes are much more profound. The
result, Eq.(1.12), is \ also remarkable in the sense that it allows us to
interpret the Veneziano amplitude from the point of view of number theory,
the theory of dynamical systems, etc. Details can be found\ in our recent
work, Ref.[5]. No such interpretation is possible to our knowledge for the
Virasoro amplitudes. For this and other reasons to be described below in
this paper, we shall consider only the Veneziano and Veneziano-like
amplitudes. \ 

In particular, now we would like to dicuss some basic analytic properties of
the 4- particle Veneziano amplitude. To this purpose we need to use the
following identities 
\begin{equation}
\sin \pi z=\pi z\prod\limits_{k=1}^{\infty }(1-(\frac{z}{k}))(1+(\frac{z}{k}%
))  \tag{1.15}
\end{equation}
and 
\begin{equation}
\frac{1}{\Gamma (z)}=ze^{-Cz}\prod\limits_{k=1}^{\infty }(1+(\frac{z}{k}%
))e^{-\dfrac{z}{k}}  \tag{1.16}
\end{equation}
with $C$ being the Euler's constant 
\begin{equation*}
C=\lim_{n\rightarrow \infty }(1+\frac{1}{2}+\frac{1}{3}+...+\frac{1}{n}-\ln
n).
\end{equation*}
\ When combined with the Veneziano condition, $\alpha (s)+\alpha (t)+\alpha
(u)=-1,$Eq$.(1.5),$ the above results allow us to write (up to a constant
factor) a typical singular portion of the Veneziano amplitude (the tachyons
are to be considered separately): 
\begin{eqnarray}
A(s,t,u) &=&\frac{1}{nlm(1-\frac{\alpha (s)}{n})}\frac{1}{(1-\frac{\alpha (t)%
}{m})}\frac{1}{(1-\frac{\alpha (u)}{l})}\times  \notag \\
&&[(1-\frac{\alpha (s)}{n})C(n)+(1-\frac{\alpha (t)}{m})C(m)+(1-\frac{\alpha
(u)}{l})C(l)]  \TCItag{1.17}
\end{eqnarray}
where $C(n)$, etc. are known constants and $m,n,l$ as some nonnegative
integers. For actual use of this result the explicit form of these constants
may be important. Looking at Eq.(1.15) we obtain, 
\begin{equation}
C(n,\alpha )=\pi \alpha \frac{1}{(1-\dfrac{\alpha }{n})}\prod\limits_{k=1}^{%
\infty }(1-(\frac{\alpha }{k}))(1+(\frac{\alpha }{k})).  \tag{1.18}
\end{equation}
where $\alpha $ can be $\alpha (s)$, etc. Clearly, this definition leads to
further simplifications, e.g. to the manifestly symmetric form:

\bigskip\ 
\begin{eqnarray}
A(s,t,u) &=&\frac{1}{nlm}[\frac{C(n,\alpha (s))}{(1-\frac{\alpha (t)}{m})}%
\frac{1}{(1-\frac{\alpha (u)}{l})}  \notag \\
&&+\frac{C(m,\alpha (t))}{(1-\frac{\alpha (s)}{n})}\frac{1}{(1-\frac{\alpha
(u)}{l})}  \notag \\
&&+\frac{C(l,\alpha (u))}{(1-\frac{\alpha (s)}{n})}\frac{1}{(1-\frac{\alpha
(t)}{m})}].  \TCItag{1.19a}
\end{eqnarray}
Consider now special case : $\alpha (s)=\alpha (t)=n.$ In this case we
obtain: 
\begin{eqnarray}
A(s &=&t,u)=\frac{1}{n^{2}m}\frac{1}{(1-\frac{\alpha (s)}{n})^{2}}%
[C(l,\alpha (u))+2C(n,\alpha (s))\frac{(1-\frac{\alpha (s)}{n})}{(1-\frac{%
\alpha (u)}{l})}  \notag \\
&=&\frac{1}{n^{2}m}\frac{1}{(1-\frac{\alpha (s)}{n})^{2}}[\sin \pi \alpha
(u)+2\sin \pi \alpha (s)]\frac{1}{(1-\frac{\alpha (u)}{l})}=0  \TCItag{1.19b}
\end{eqnarray}
This result is in accord with that of Ref.[6\textbf{]} where it was obtained
differently. The tachyonic case is rather easy to consider now. Indeed,
using Eq.s(1.6), (1.15),(1.16) and taking into account the Veneziano
condition, let us assume that, say, that $\alpha (s)=0$ then, in view of
symmetry of Eq.s.(1.19a) and (1.19b), we need to let $\alpha (t)=0$ as well
to check if Eq(1.19b) holds. This leaves us with the option: $\alpha (u)=-1.$%
With such a constraint we obtain $($ since $\Gamma (1)=1),$%
\begin{equation*}
A(s,t,u)=\frac{\pi }{\alpha (t)}+\frac{\pi }{\alpha (s)}-\frac{\pi }{\alpha
(t)}-\frac{\pi }{\alpha (s)}=0,
\end{equation*}
as required. Hence, indeed, even in the tachyonic case Eq.(1.19b) holds in
accord with earlier results [6]. This means that one cannot observe the
tachyons in both channels simultaneously. But even to observe them in one
channel is unphysical. Moreover, Eq.(1.19b) implies that only situations for
which $\alpha (s)\neq \alpha (t)\neq \alpha (u)$ can be in principle
physically observable.

By combining the Veneziano condition with such a constraint leaves us with
the following options:

a) $\alpha (s),\alpha (t)>0,\alpha (u)<0$;

b) $\alpha (s)>0,\alpha (t),\alpha (u)<0$ plus the rest of cyclically
permuted inequalities.

This means that not only the tachyons of the type $\alpha (s)=0$ (or $\alpha
(t)=0$ , or $\alpha (u)=0$ ) could be present but also those for which, for
example, $\alpha (s)<0.$ This is so because according to known results for
standard open string theory [2] in 26 space-time dimensions $\alpha (s)=1+%
\frac{1}{2}s$. When $\alpha (s)=0$ such convention produces the only one
tachyon: $s=-2$ $=M^{2}$ and the whole spectrum (open string) is given by $%
M^{2}=-2,0,2,...,2n$ where $n$ is a non negative integer. Incidentally, for
the closed bosonic string under the same conditions the spectrum is known to
be $M^{2}=-8,0,8,...,8n$. No other masses are permitted. \ The requirements
like those in a) and b) above produce additional complications however. For
instance, let $\alpha (s)=1$ and consider the following option : $\alpha
(t)=3,$ so that we should have $\alpha (u)=-5.$ This leads us to the tachyon
mass : $M^{2}$ $=-12$ . It is absent in the spectrum of both types of
bosonic strings. The emerging \ apparent difficulty can actually be bypassed
somehow due to the following chain of arguments.

\textbf{Remark} \textbf{1.2}.Consider, for example, the amplitude $V(s,t)$
and let both $s$ and $t$ be non tachyonic and, of course, $\alpha (s)\neq
\alpha (t)$ . Then, naturally, $\alpha (u)<0$ \textit{is} tachyonic but,
when we use Eq.(1.19 a)), we notice at once that this creates no difficulty
since $\alpha <0$ condition simply will eliminate the resonance in the
respective channels, e.g. if $V(s,t)$ will have poles for both $s$ and $t$ 
\textit{then} \textit{the same particle resonances will occur} in $V(s,u)$
and $V(t,u)$ channels so that, \textit{except the case} $\alpha (s)=0$
leading to the pole with mass $M^{2}=-2,$ no other tachyonic states will
show up as resonances and, hence, they cannot be observed. This argument is
important for designing of the bosonic string model but is in apparent
violation of the Veneziano condition, Eq.(1.5). It is violated if the
tachyons of larger negative mass are \textit{not} present in the spectrum.
Since the Veneziano condition is caused by the energy-momentum conservation,
it cannot be readily replaced by something else. The arguments just
presented explain in part the inadequacy of the existing formulation of the
model reproducing the Veneziano amplitudes.

At the same time, irrespective to the hypothetical model one uses for
reproduction of these amplitudes, based on the arguments just presented it
should be clear that, \textit{effectively}, we have only two possibilities
for resonances to be observed. That is experimentally \ (in view of the
Veneziano condition) we can either observe the resonances for combinations $%
\mathcal{V}_{u}(s)$ or $\mathcal{V}_{u}(t)$. Clearly, $\mathcal{V}_{u}(s)=$ $%
V(s,t)+$ $V(s,u)$ and $\mathcal{V}_{u}(t)=V(t,s)+V(t,u)$. Such a conclusion
is valid only if we require $V(s,t)=V(t,s)$, etc. It is surely the case for
the Veneziano amplitude, Eq.(1.19a). Accordingly, should the Veneziano
amplitude be free of tachyons (e.g. $\alpha (s)=0),$ it would be perfectly
acceptable. In the light of the results just obtained it can be effectively
written as 
\begin{equation}
A(s,t,u)=\mathcal{V}_{u}(s)+\mathcal{V}_{u}(t).  \tag{1.20}
\end{equation}
The result, Eq.(1.20), survives when, instead of the Veneziano, the
tachyon-free Veneziano-like amplitudes are used. These are discussed in
Section 3. \ Mathematical arguments leading to the construction of models
associated with such amplitudes are discussed in detail in Parts 2-4 of this
work.

To complete this section, we would like to provide a brief preview of
arguments leading to reconstruction of these models. Clearly, such a preview
is only a small part of other arguments to be discussed. We believe, that
the arguments presented below should be especially appealing to readers
familiar with string theory.

\bigskip\ Following Hirzebruch and Zagier [7] \ let us consider an identity 
\begin{eqnarray}
\frac{1}{(1-tz_{0})\cdot \cdot \cdot (1-tz_{k})} &=&(1+tz_{0}+\left(
tz_{0}\right) ^{2}+...)\cdot \cdot \cdot (1+tz_{n}+\left( tz_{n}\right)
^{2}+...)  \notag \\
&=&\sum\limits_{n=0}^{\infty
}(\sum\limits_{k_{0}+...+k_{k}=n}z_{0}^{k_{0}}\cdot \cdot \cdot
z_{k}^{k_{k}})t^{n}.  \TCItag{1.21}
\end{eqnarray}%
When $z_{0}=...=z_{n}=1,$the inner sum in the last expression provides the
total number of monomials of the type $z_{0}^{k_{0}}\cdot \cdot \cdot
z_{n}^{k_{n}}$ with $k_{0}+...+k_{k}=n$. The total number of such monomials
is given by the binomial coefficient [8] 
\begin{equation}
p(k,n)\equiv \left( 
\begin{array}{c}
n+k \\ 
k%
\end{array}%
\right) =\frac{(n+1)(n+2)\cdot \cdot \cdot (n+k)}{k!}.  \tag{1.22}
\end{equation}%
For this special case Eq.(1.21) is converted to a useful expansion, 
\begin{equation}
P(k,t)\equiv \frac{1}{\left( 1-t\right) ^{k+1}}=\sum\limits_{n=0}^{\infty
}p(k,n)t^{n}.  \tag{1.23}
\end{equation}%
In view of the integral representation of beta function given by Eq.(1.2), \
we replace $k+1$ by $\alpha (s)+1$ in Eq.(1.23) and use it in the beta
function representation of $V(s,t)$ given by Eq.(1.2). Straightforward
calculation produces the following known in string theory \textbf{[}2\textbf{%
]} result: 
\begin{equation}
V(s,t)=-\sum\limits_{n=0}^{\infty }p(\alpha (s),n)\frac{1}{\alpha (t)-n}. 
\tag{1.24}
\end{equation}%
The r.h.s. of Eq.(1.24) can be interpreted as the Laplace transform of the
partition function, Eq.(1.23). Such an interpretation is not merely formal.
To see this, following Vergne, Ref.[9], let us consider a region $\Delta _{k}
$ of $\mathbf{R}^{k}$ consisting of all points $\nu =(t_{1},t_{2},...,t_{k}),
$ such that coordinates $t_{i}$ of $\nu $ are non negative and satisfy the
constraint: $t_{1}+t_{2}+...+t_{k}\leq 1.$ Clearly, such a restriction is
characteristic for the simplex in \textbf{R}$^{k}.$ Consider now a \textit{%
dilated} simplex: $n\Delta _{k}$ for some nonnegative integer $n$. The
volume of $n\Delta _{k}$ is easily calculated and is known to be 
\begin{equation}
\text{vol(}n\Delta _{k})=\frac{n^{k}}{k!}.  \tag{1.25}
\end{equation}%
Next, let us consider points $\nu =(u_{1},u_{2},...,u_{k})$ with \textit{%
integral} coordinates \textit{inside} the dilated simplex $n\Delta _{k}.$
The total number of points with integral coordinates inside $n\Delta _{k}$
is given by $p(k,n)$, Eq.(1.22), i.e. 
\begin{equation}
p(k,n)=\left\vert n\Delta _{k}\cap \mathbf{Z}^{k}\right\vert =\frac{%
(n+1)(n+2)\cdot \cdot \cdot (n+k)}{k!}.  \tag{1.26}
\end{equation}%
The function $p(k,n)$ happen to be the non negative integer. It arises
naturally as the dimension of the quantum Hilbert space associated (through
the coadjoint orbit method) with the symplectic manifold of dimension $2k$
constructed by \textquotedblright inflating\textquotedblright\ $\Delta _{k}.$
Although the details \ related to such symplectic manifods and the
associated with them dynamical systems will be provided in Parts 2 and 3,
the next section supplies additional relevant information.

\bigskip

\section{Relationship between the hypergeometric functions and the Veneziano
amplitudes}

\bigskip

The fact that the hypergeometric functions are the simplest solutions of the
Knizhnik-Zamolodchikov equations of CFT is well documented [10]. The
connection between these functions and the toric varieties (to be discussed
in Part 2) had been also developed in papers by Gelfand, Kapranov and
Zelevinsky (GKZ) [11]. Hence, we see no need in duplication of their results
in this work. Instead, we would like to discuss other aspects of
hypergeometric functions and their connections with the Veneziano amplitudes
emphasizing similarities and differences between strings and CFT.

For reader's convenience, we begin by introducing some standard notations.
In particular, let 
\begin{equation*}
(a,n)=a(a+1)(a+2)\cdot \cdot \cdot (a+n-1)
\end{equation*}
and, more generally, $(a)=(a_{1},...,a_{p})$ and $(c)=(c_{1},...,c_{q}).$
With help of these notations the $p,q$-type hypergeometric function can be
written as 
\begin{equation}
_{p}F_{q}[(a);(c);x]=\sum\limits_{n=0}^{\infty }\frac{(a_{1,}n)\cdot \cdot
\cdot (a_{p},n)}{(c_{1,}n)\cdot \cdot \cdot (c_{q},n)}\frac{x^{n}}{n!}. 
\tag{2.1}
\end{equation}
In particular, the hypergeometric function in the form known to Gauss is
just $_{2}F_{1\text{ }}=$ $F[a,b;c;x]$. Practically all elementary functions
and almost all special functions can be obtained as special cases of the
hypergeometric function just defined [12].

We are interested \ in connections between the hypergeometric functions and
the Schwarz-Christoffel \ (S-C) mapping problem. The essence of this problem
lies in finding a function $\varphi (\zeta )=z$ which maps the upper half
plane $\func{Im}\zeta >0$ (or, equivalently, the unit circle) into the
exterior of the $n-$sided polygon located on the Riemann sphere considered
as one dimensional complex projective space $\mathbf{CP}^{1}$ (i.e. $z\in 
\mathbf{CP}^{1}).$ Traditionally, the pre images $a_{1},...,a_{n}$ of the\
polygon vertices \ located at points $b_{1},...,b_{n}$ in $\mathbf{CP}^{1}$
are placed onto $x$ axis of $\zeta $-plane so that $\varphi (a_{i})=b_{i}$ , 
$i=1-n$. Let the interior angles of the polygon be $\pi \alpha _{1},...,\pi
\alpha _{n\text{ }}$respectively. Then the exterior angles $\mu _{i}$\ are
defined trough relations $\pi \alpha _{i}+\pi \mu _{i}=\pi $ , $i=1-n.$ The
exterior angles are subject to the constraint: $\sum\nolimits_{i=1}^{n}\mu
_{i}=2.$ The above data \ allow us to write for the S-C mapping function the
following \ known expression: 
\begin{equation}
\varphi (\zeta )=C\int\limits_{0}^{\zeta }(t-a_{1})^{-\mu _{1}}\cdot \cdot
\cdot (t-a_{n})^{-\mu _{n}}+C^{\prime }.  \tag{2.2}
\end{equation}%
If one of the points, say $a_{n},$ is located at infinity, it can be shown
that in the resulting formula for mapping the last term under integral can
be deleted.

Consider now the simplest but \ relevant example of mapping of the upper
half plane into triangle with angles $\alpha ,\beta $ and $\gamma $ subject
to Euclidean constraint: $\alpha +\beta +\gamma =1.$ Let, furthermore, $%
a_{1}=0,a_{2}=1$ and $a_{3}=\infty $. Using Eq.(2.2) (with $C=1$) we obtain
for the length $c$ of the side of the triangle: 
\begin{equation}
c=\int\limits_{0}^{1}\left| \frac{d\varphi (\zeta )}{d\zeta }d\zeta \right|
=\int\limits_{0}^{1}z^{\alpha -1}(1-z)^{\beta -1}=B(\alpha ,\beta )=\frac{%
\Gamma (\alpha )\Gamma (\beta )}{\Gamma (1-\gamma )}.  \tag{2.3}
\end{equation}
Naturally, two other sides can be determined the same way. Much more
efficient \ is to use the \ familiar elementary trigonometry relation 
\begin{equation*}
\frac{c}{\sin \pi \gamma }=\frac{b}{\sin \pi \beta }=\frac{a}{\sin \pi
\alpha }.
\end{equation*}
Then, using Eq.(1.9a), we obtain for the sides the following results : $c=%
\frac{1}{\pi }[\sin \pi \gamma ]\Gamma (\alpha )\Gamma (\beta )\Gamma
(\gamma );$ $b=\frac{1}{\pi }[\sin \pi \beta ]\Gamma (\alpha )\Gamma (\beta
)\Gamma (\gamma );$ $a=\frac{1}{\pi }[\sin \pi \alpha ]\Gamma (\alpha
)\Gamma (\beta )\Gamma (\gamma ).$ The perimeter length $\mathcal{L}=a+b+c$
\ of the triangle is just the full Veneziano amplitude, Eq.(1.1). As is well
known [13], the conformal mapping with \ Euclidean constraint $\alpha +\beta
+\gamma =1$ can be performed only for $3$ sets of \textit{fixed} angles.
Another 4 sets of angles belong to the spherical case: $\alpha +\beta
+\gamma >1,$ while countable infinity of angle sets exist for the hyperbolic
case: $\alpha +\beta +\gamma <1.$ Hence, the associated with such mappings
Fuchsian-type equations used in some string theory formulations will not be
helpful in deriving the Veneziano amplitudes. \ These equations are useful
however in the CFT as is well known [10].

It is well documented that, to some extent, development of the bosonic
string theory is unseparable from attempts at multidimensional
generalization of Euler's beta function [6]. Analogous developments also
took place in the theory of hypergeometric functions where they proceeded
along two related lines. To illustrate the key ideas, following Mostow and \
Deligne [14], \ let us consider the \ standard hypergeometric function
which, up to a constant\footnote{%
To indicate this we use symbol $\dot{=}$.}, is given by 
\begin{equation}
F[a,b;c;x]\dot{=}\int\limits_{1}^{\infty }u^{a-c}(u-1)^{c-b-1}(u-x)^{-a}du. 
\tag{2.4}
\end{equation}
The multidimensional (multivariable) analogue of the above function \
according to Picard (in notations of Mostow and Deligne) is given by 
\begin{equation}
F[x_{2},...,x_{n+1}]=\int\limits_{1}^{\infty }u^{-\mu _{0}}(u-1)^{-\mu
_{1}}\prod\limits_{i=2}^{n+1}(u-x_{i})^{-\mu _{i}}du,  \tag{2.5}
\end{equation}
provided that $x_{0}=0,x_{1}=1$ and, as before, $\sum\nolimits_{i=0}^{n}\mu
_{i}=2.$ At the same time, using the alternative representation of $%
F[a,b;c;x]$ given by 
\begin{equation}
F[a,b;c;x]\dot{=}\int\limits_{0}^{1}z^{b-1}(1-z)^{c-b-1}(1-zx)^{-a}dz 
\tag{2.6}
\end{equation}
one obtains as well the following multidimensional generalization: 
\begin{equation}
F[\alpha ,\beta ,\beta ^{\prime },\gamma ;x,y]\dot{=}\int \int\limits 
_{\substack{ u\geq 0,v\geq 0  \\ u+v\leq 1}}u^{\beta -1}v^{\beta ^{\prime
}-1}(1-u-v)^{\gamma -\beta -\beta ^{\prime }-1}(1-ux)^{-\alpha }(1-v\unit{y}%
)^{-\alpha ^{\prime }}dudv  \tag{2.7}
\end{equation}
This result was obtained by Horn already at the end of 19th century and \
was subsequently reanalyzed and extended by GKZ. Looking at the last
expression one can design \ by analogy the multidimensional extension of the
Euler's beta function. In view of Eq.(1.2), it is given by the following
integral attributed to Dirichlet: 
\begin{equation}
\mathcal{D}(x_{1},...,x_{k})=\int \int\limits_{\substack{ u_{1}\geq
0,...,u_{k}\geq 0  \\ u_{1}\text{ }+\cdot \cdot \cdot +u_{k}\leq 1}}\unit{u}%
_{1}^{x_{1}-1}\unit{u}_{2}^{x_{2}-1}...\unit{u}%
_{k}^{x_{k}-1}(1-u_{1}-...-u_{k})^{x_{k+1}-1}du_{1}...du_{k}.  \tag{2.8}
\end{equation}

In this integral let $t=u_{1}+...+u_{k}$. This allows us to use already
familiar expansion, Eq.(1.23). In addition, however, we would like to use
the following identity 
\begin{equation}
t^{n}=(u_{1}+...+u_{k})^{n}=\sum\limits_{n=(n_{1},...,n_{k})}\frac{n!}{%
n_{1}!n_{2}!...n_{k}!}u_{1}^{n_{1}}\cdot \cdot \cdot u_{k}^{n_{k}}  \tag{2.9}
\end{equation}%
with restriction $n=n_{1}+...+n_{k}.$ This type of identity was used earlier
in our work on Kontsevich-Witten model [15]. Moreover, from the same paper
we obtain the alternative and very useful form of the above expansion 
\begin{equation}
(u_{1}+...+u_{k})^{n}=\sum\limits_{\lambda \vdash k}f^{\lambda }S_{\lambda
}(u_{1},...,u_{k})  \tag{2.10}
\end{equation}%
where the Schur polynomial $S_{\lambda }$ is defined by 
\begin{equation}
S_{\lambda }(u_{1},...,u_{k})=\sum\limits_{n=(n_{1},...,n_{k})}K_{\lambda
,n}u_{1}^{n_{1}}\cdot \cdot \cdot u_{k}^{n_{k}}  \tag{2.11}
\end{equation}%
with coefficients $K_{\lambda ,n}$ known as Kostka numbers [16], $\
f^{\lambda }$ being the number of standard Young tableaux of shape $\lambda $
and the notation $\lambda \vdash k$ meaning that $\lambda $ is partition of $%
k$. Through such connection with Schur polynomials one can develop
connections with Kadomtsev-Petviashvili (KP) hierarchy of nonlinear exactly
integrable systems on one hand\ and with the theory of Schubert varieties on
another [15\textbf{]}. We shall provide more details \ on such a connection
in Part 4. 

Use of Eq.(2.9) in Eq.(2.8) produces, after effectively performing the
multiple Laplace transform, the following part of the multiparticle
Veneziano amplitude 
\begin{equation}
A(1,...k)=\frac{\Gamma _{n_{1}...n_{k}}(\alpha (s_{k+1}))}{(\alpha
(s_{1})-n_{1})\cdot \cdot \cdot (\alpha (s_{k})-n_{k})}.  \tag{2.12}
\end{equation}%
Even though the residue $\Gamma _{n_{1}...n_{k}}(\alpha (s_{k+1}))$ contains
all the combinatorial factors, the obtained result should still be
symmetrized (in accord with the 4-particle case considered by Veneziano) in
order to obtain the full murtiparticle Veneziano amplitude. Since in such
general multiparticle case the same expansion, Eq.(1.23), was used,
arguments of the previous section can be applied to the present case as well
thus leading to the same model considered by Vergne, Ref.[9]. Details will
be discussed in Parts 2 and 3.

\bigskip\ 

\section{Veneziano amplitudes from Fermat hypersurfaces}

\subsection{General considerations}

\bigskip

In 1967-a year earlier than Veneziano's paper was published, the paper [17]
by Chowla ans Selberg had appeared relating Euler's beta function to the
periods of elliptic integrals. The result by Chowla and Selberg was
generalized by Andre Weil whose two influential papers [18,19] have brought
into picture the periods of Jacobians of the Abelian varieties, Hodge rings,
etc. Being motivated by these papers, Benedict Gross had writtern a paper
[20] in which the beta function appears as period associated with the
differential form ''living'' on the Jacobian of the Fermat curve. His
results as well as those by Rohrlich (placed in the appendix to Gross paper)
have been subsequently documented in the book by Lang [21]. Although in the
paper by Gross, Ref.[20], the multidimensional extension of beta function is
briefly considered, e.g. read page 207 of Ref.[20], the computational
details were not provided hovewer. We provide these details below following
some ideas developed in lecture notes by Deligne [22]. To obtain the
multidimensional extension of beta function following logic of paper by
Gross, one needs to replace the Fermat curve by the Fermat hypersufrace, to
embed it into projective space and by complexification treat it as K\"{a}%
hler manifold. The differential forms living on such manifold are associated
with periods of Fermat hypersurface. In Part 2 and 3 of this work we shall
argue that thus obtained K\"{a}hler manifold is of Hodge type. We also will
provide arguments independent from those by Weil [18,19] needed to arrive at
the same conclusions. In his lecture notes Deligne noticed that the Hodge
theory requires some essential changes (e.g. mixed Hodge structures, etc.)
in the case if the Hodge-K\"{a}hler manifolds possess singularities. Such
modifications may be needed upon development of the formalism we are about
to discuss. A monograph by Carlson et al, Ref.[23], contains an up to date
exaustive information regarding such modifications, etc. Fortunately, to
obtain the multiparticle Veneziano amplitudes such complications are not
nesessary. Hence we proceed directly with description of main ideas.

To illustrate these ideas, in accordrd with Ref.[23], we are following the
arguments by Griffiths [24]. To this purpose, let us begin with the simplest
example of calculation of the following period integral 
\begin{equation}
\pi (\lambda )=\oint_{\Gamma }\frac{dz}{z(z-\lambda )}  \tag{3.1}
\end{equation}
taken along the closed contour $\Gamma $ in the complex $z$-plane. Since
this integral depends upon parameter $\lambda $ the period $\pi (\lambda )$
is some function of $\lambda .$ It can be determined by straightforward
differentiation of $\pi (\lambda )$ with respect to $\lambda $ thus leading
to the desired differential equation 
\begin{equation}
\lambda \pi ^{\prime }(\lambda )+\pi (\lambda )=0  \tag{3.2}
\end{equation}
enabling us to calculate $\pi (\lambda ).$ This simple result can be vastly
generalized to cover the case of period integrals of the type 
\begin{equation}
\Pi (\lambda )=\oint\limits_{\Gamma }\frac{P\left( z_{1},...,z_{n}\right) }{%
Q\left( z_{1},...,z_{n}\right) }dz_{1}\wedge dz_{2}\cdot \cdot \cdot \wedge
dz_{n}.  \tag{3.3}
\end{equation}
The equation $Q\left( z_{1},...,z_{n}\right) =0$ determines algebraic
variety. It may conatain a parameter (or parameters) $\lambda $ so that the
polar locus of values of $z^{\prime }s$ satisfying equation $Q=0$ depends
upon this parameter(s). By analogy with Eq.(3.2), it is possible to obtain a
set of differential equations of P-F type. This was demonstrated originally
by Manin [25]. In this work we are not going to develop this line of
research however. Instead, following Griffiths \textbf{[}24], we want to
analyze in some detail the nature of the expression under the sign of
integral in Eq.(3.3).

If $\mathbf{x}=(x_{0,}...,x_{n})$ are homogenous coordinates of a point in 
\textit{projective space}\textbf{\ }and $z=(z_{1},...,z_{n})$ are the
associated coordinates of the point in the \textit{affine space} where $%
z_{i}=x_{i}/x_{0}$ , then the rational $n-$form $\omega $ is given in the 
\textit{affine} space by 
\begin{equation}
\omega =\frac{P\left( z_{1},...,z_{n}\right) }{Q\left(
z_{1},...,z_{n}\right) }dz_{1}\wedge dz_{2}\cdot \cdot \cdot \wedge dz_{n} 
\tag{3.4}
\end{equation}
with rational function $P/Q$ being a quotient of two homogenous polynomials
of the \textit{same} degree. Upon substitution: $z_{i}=x_{i}/x_{0}$, the
form $dz_{1}\wedge dz_{2}\cdot \cdot \cdot \wedge dz_{n}$ changes to 
\begin{equation*}
dz_{1}\wedge dz_{2}\cdot \cdot \cdot \wedge dz_{n}=\left( x_{0}\right)
^{-(n+1)}\sum\limits_{i=0}^{n}(-1)^{i}x_{i}dx_{0}\wedge ...\wedge d\hat{x}%
_{i}\wedge ...\wedge dx_{n}
\end{equation*}
where the hat on the top of $x_{i}$ means that it is excluded from the
product. It is convenient now to define the form $\omega _{0}$ via 
\begin{equation*}
\omega _{0}:=\sum\limits_{i=0}^{n}(-1)^{i}x_{i}dx_{0}\wedge ...\wedge d\hat{x%
}_{i}\wedge ...\wedge dx_{n}
\end{equation*}
so that in terms of \textit{projective space} coordinates the form $\omega $
can be rewritten as 
\begin{equation}
\omega =\frac{p(\mathbf{x})}{q(\mathbf{x})}\omega _{0}  \tag{3.5}
\end{equation}
where $p(\mathbf{x})=P(\mathbf{x})$ and $q(\mathbf{x})=Q(\mathbf{x}%
)x_{0}^{n+1}$ or, in more symmetric form, $q(\mathbf{x})=Q(\mathbf{x}%
)x_{0}\cdot \cdot \cdot x_{n}.$ In this case the degree of denominator of
the rational function $p/q$ is that of numerator $+(n+1)$. This is the
result of the Corollary 2.11 of Griffith's paper [24]. Conversely, each
homogenous differential form $\omega $ in projective space can be written in
affine space upon substitution : $x_{0}=1$ and $x_{i}=z_{i},i\neq 0.$

\ We would like to take advantage of this fact now. To this purpose, as an
example, we would like to study the period integrals associated with
equation describing Fermat hypersurface in complex projective space 
\begin{equation}
\mathcal{F(}N):x_{0}^{N}+\cdot \cdot \cdot +x_{n}^{N}+x_{n+1}^{N}=0. 
\tag{3.6}
\end{equation}
We would like to consider the set of independent linear forms $%
x_{i}^{<c_{i}>}$ , $i=0-(n+1),$where $<c_{i}>$ denotes representative of $%
c_{i}$ in \textbf{Z} such that for now $1\leq <c_{i}>\leq N-1\footnote{%
These limits for $<c_{i}>$ are in accord with Gross [20], page 198.
Subsequently, they will be changed below to $1\leq <c_{i}>\leq N$.}$.They
can be interpreted as\ the set of \ hyperplanes in $\mathbf{C}^{n+1}$ whose
complement is complex algebraic torus $T$ as it will be explained in detail
in Part 2. We want to consider the form $\omega $ living at the intersection
of $T$ with $\mathcal{F}$. To this purpose it is convenient to introduce the
average $<c>$ as follows 
\begin{equation}
<c>=\frac{1}{N}\sum\limits_{i}<c_{i}>.  \tag{3.7}
\end{equation}
The numbers $c_{i}$ belong to the set $X(S^{1})$ given by 
\begin{equation}
X(S^{1})=\{\bar{c}\in (\mathbf{Z}/N\mathbf{Z})^{n+2}\equiv (\mathbf{Z}/N%
\mathbf{Z})\times \cdot \cdot \cdot \times (\mathbf{Z}/N\mathbf{Z})\mid \bar{%
c}=(c_{0},...,c_{n+1}),\tsum\limits_{i}c_{i}=0\func{mod}N\}  \tag{3.8}
\end{equation}
The true meaning of the condition $\tsum\limits_{i}c_{i}=0\func{mod}N$ is
illustrated \ below by using the Fermat hypersurface $\mathcal{F(}N\mathcal{)%
}$ as an example. In this case the form $\omega ,$Eq$.(3.5),$ is given by 
\begin{equation}
\omega =\frac{x_{0}^{<c_{0}>-1}\cdot \cdot \cdot x_{n+1}^{<c_{n+1}>-1}}{%
\left( x_{0}^{N}+\cdot \cdot \cdot +x_{n}^{N}+x_{n+1}^{N}\right) ^{<c>}}%
\text{ }\omega _{o}.  \tag{3.9}
\end{equation}
By design, it satisfies all of the requirements of the Corollary 2.11 of
Griffith's paper.

\subsection{The 4-particle Veneziano-like amplitude}

\bigskip

Using Eq.(3.9), let us consider the simplest \ but important case : $n=1.$
It is relevant for calculation of 4-particle Veneziano-like amplitude.
Converting $\omega $ into affine form according to Griffiths prescription we
obtain the following result for the period integral: 
\begin{equation}
I_{aff}=\oint\limits_{\Gamma }\frac{1}{\left( x_{1}^{N}+x_{2}^{N}\mp
1\right) }dx_{1}^{<c_{1}>}\wedge dx_{2}^{<c_{2}>}.  \tag{3.10a}
\end{equation}
The $\pm $ sign in the denominator requires some explanation. Indeed, let us
for a moment restore the projective form of this integral. By doing so,we
obtain the following integral: 
\begin{equation}
I_{proj}=\oint\limits_{\Gamma }\frac{%
z_{1}^{<c_{1}>}z_{2}^{<c_{2}>}z_{0}^{<c_{0}>}}{\left( z_{1}^{N}+z_{2}^{N}\pm
z_{0}^{N}\right) }(\frac{dz_{1}^{{}}}{z_{1}}\wedge \frac{dz_{2}^{{}}}{z_{2}}-%
\frac{dz_{0}^{{}}}{z_{0}}\wedge \frac{dz_{2}^{{}}}{z_{2}}+\frac{dz_{0}^{{}}}{%
z_{0}}\wedge \frac{dz_{1}^{{}}}{z_{1}}).  \tag{3.10b}
\end{equation}
It is manifestly symmetric with respect to permutation of its arguments by
construction. In addition, we would like it to be invariant with respect to
scale transformations of the type : $z_{j}$\ $\rightarrow z_{j}\xi ^{j}$ ,\
where $\xi ^{j}=\exp (\pm i\frac{2\pi j}{N})$ with $1\leq j\leq N-1.$ Such
scaling is used extensively in the theory of invariants of the
pseudo-reflection groups. Its meaning will be discussed in Part 2 in
connection with invariance properties of the Veneziano and Veneziano-like
amplitudes. For now, it is sufficient to realize only that the numerator of
the integrand in Eq.(3.10b) as a whole acquires the following phase factor: $%
\exp \{i\frac{2\pi }{N}(<c_{1}>j+<c_{2}>k+<c_{0}>l)\}$. Since by design the
integral \ $I_{proj}$ is made to satisfy the Corollary 2.11 discussed in
previous subsection, it is sufficient to require 
\begin{equation}
<c_{1}>j+<c_{2}>k+<c_{0}>l=N  \tag{3.11a}
\end{equation}
in order to make it manifestly scale invariant (torus action invariant in
the terminology of Part 2). \ We shall call Eq.(3.11a) the ''Veneziano
condition'' while Eq.(3.11b) (below) we shall call the ''Shapiro-Virasoro''
condition.\footnote{%
These names are given by analogy with the existing terminilogy for the open
(Veneziano) and closed (Shapiro-Virasoro) bosonic strings. Clearly, in the
present context they emerge for reasons different from those used in
conventional formulations.}. Transition from the projective to affine space
breaks the permutational symmetry firstly because of selecting, say, $z_{0}$ 
$($and requiring it to be one) and, secondly, by possibly switching the sign
in front of $z_{0}.$ The permutational symmetry can be restored in the style
of Veneziano, e.g. see Eq.(1.1). The problem of switching the sign in front
of $z_{0}$ can be treated similarly but requires extra care. This is so
because instead of the factor $\xi ^{j}$ used above we could have used $%
\varepsilon ^{j}$, where $\varepsilon =\exp (\pm i\dfrac{\pi }{N})$\footnote{%
Both options will be explained in Part 2 from the point of view of the
concept of the torus action. For alternative point of view, please read
Ref.[26].} Use of such a factor makes the integral $I_{proj}$ also torus
action invariant. But for this case the condition, Eq.(3.11a), has to be
changed into 
\begin{equation}
<c_{1}>j+<c_{2}>k+<c_{0}>l=2N  \tag{3.11b}
\end{equation}
in accord with Lemma 1 of Gross [20]. By such a change we are in apparent
disagreement with the Corollary 2.11 by Griffiths. We write ''apparent''
because, fortunately, there is a way to reconcile the Corollary 2.11. by
Griffiths with Lemma 1 by Gross. It will be discussed below. Already
assuming that this is the case, we notice that there are at least two
different classes of transformations leaving $I_{proj}$ unchanged. When
switching to the affine form these two classes are not equivalent: the first
leads to differential forms of the first kind while the second-to the second
kind [20,21]. Both are living on the Jacobian variety $J(N)$ associated with
the Fermat surface $\mathcal{F(}N)$ : $z_{1}^{N}+z_{2}^{N}\pm 1=0.$ It
happens, that physically more relevant are the forms of the second kind. We
would like to describe them now.

We begin by noticing that in switching from the projective to affine space
the following set of $3N$ points (at infinity) should be deleted from the
Fermat curve $z_{1}^{N}+z_{2}^{N}+z_{3}^{N}$ $=0$. These are: $(\varepsilon
\xi ^{j},0,1),$ $(0,\varepsilon \xi ^{j},1),(\varepsilon ^{2}\xi
^{j},\varepsilon \xi ^{j},0)$ respectively [26]. By assuming that this is
the case and paramerizing $z_{1}$ and $z_{2\text{ \ }}$as $z_{1}=$ $%
\varepsilon t_{1}^{\frac{1}{N}}$ and $z_{2}=\varepsilon t_{2}^{\frac{1}{N}}$%
, we obtain the simplex equation $t_{1}+t_{2}=1$ as deformation retract for $%
\mathcal{F(}N)\footnote{%
The rationale for such substitutions is explained in Part 2.}.$ After this,
Eq.(3.10a) acquires the following form : 
\begin{equation}
I_{aff}=\xi ^{j<c_{1}>+k<c_{2}>}\frac{1}{N^{2}}\oint\limits_{\Gamma }\frac{%
\varepsilon ^{<c_{1}>}t_{1}^{\frac{<c_{1}>}{N}}\varepsilon ^{<c_{2}>}t_{2}^{%
\frac{<c_{2}>}{N}}}{\left( t_{1}+t_{2}-1\right) }\frac{dt_{1}}{t_{1}}\wedge 
\frac{dt_{2}}{t_{2}}.  \tag{3.12}
\end{equation}
The overall phase factor guarantees the linear independence of the above
period integrals [21] in view of the well-known result: $1+\xi ^{r}+\xi
^{2r}+...+\xi ^{(N-1)r}=0$ . It will be omitted for brevity in the rest of
our discussion.

To calculate $I_{aff}$ we need to use generalization of the method of
residues for multidimensional complex integrals as developed by Leray [27]
and discussed in physical context by Hwa and Teplitz [28] and others. From
this reference we find that taking the residue can be achieved either by
dividing the differential form in Eq.(3.12) by $ds=t_{1}dt_{1}+t_{2}dt_{2}$
or, what is equivalent, by writing instead of Eq.(3.12) the following
physically suggestive result 
\begin{equation}
I_{aff}=\frac{1}{N^{2}}\oint\limits_{\Gamma }\varepsilon ^{<c_{1}>}t_{1}^{%
\frac{<c_{1}>}{N}}\varepsilon ^{<c_{2}>}t_{2}^{\frac{<c_{2}>}{N}}\frac{dt_{1}%
}{t_{1}}\wedge \frac{dt_{2}}{t_{2}}\delta (t_{1}+t_{2}-1)  \tag{3.13}
\end{equation}
to be discussed further in Part 3. For the time being, taking into account
that $t_{2}=1-t_{1}$,\ after calculating the Leray residue we obtain, 
\begin{equation}
I_{aff}=\frac{1}{N^{2}}\int\limits_{0}^{1}\unit{u}^{\frac{<c_{1}>}{N}-1}(1-%
\text{u})^{\frac{<c_{2}>}{N}-1}d\text{u}=\frac{1}{N^{2}}B(a,b),  \tag{3.14}
\end{equation}
where $B(a,b)$ is Euler's beta function (as in Eq.(1.2))\ with $a=\frac{%
<c_{1}>}{N}$ and $b=\frac{<c_{2}>}{N}.$ The phase factors had been \
temporarily suppressed for the sake of comparison with the results of
Rohrlich \ [20] (published as an Appendix to paper by Gross and also
discussed in the book by\ Lang [21]). To make such comparison, we need to
take into account the multivaluedness of the integrand above if it is
considered in the standard complex plane. Referring our readers to Ch-r 5 of
Lang's book [21] allows us to avoid rather long discussion about the
available choices of integration contours. Proceeding in complete analogy
with the case considered by Lang, we obtain the period $\Omega (a,b)$ of the
differential form $\omega _{a,b}$ of the \textit{second} kind living on $%
J(N):$ 
\begin{equation}
\frac{\Omega (a,b)}{N}=\frac{1}{N}\oint\limits_{\Gamma }\omega _{a,b}=\frac{1%
}{N^{2}}(1-\varepsilon ^{<c_{1}>})(1-\varepsilon ^{<c_{2}>})B(a,b). 
\tag{3.15}
\end{equation}
The Jacobian $J(N)$ is related to the Fermat curve $\mathcal{F(}N)$
considered as Riemann surface of genus $g=\frac{1}{2}(N-1)(N-2).$\ Obtained
result differs from that by Rohrlich only by phase factors : $\varepsilon
^{\prime }s$ instead of $\xi ^{\prime }s.$ The number of such periods is
determined by the inequalities of the type $1\leq <c_{i}>\leq N-1.$ In
addition to the differential forms of the second kind there are also the
differential forms of the \textit{third} kind living on $\mathcal{F(}N).$
They can be easily obtained from that of the second kind by relaxing the
condition $1\leq <c_{i}>\leq N-1$ to $1\leq <c_{i}>\leq N$, Lang [21], page
39. The differential forms of the second kind are associated with the de
Rham cohomology classes H$_{DR}^{1}($ $\mathcal{F(}N),\mathbf{C})$ (Gross
[20], Lemma 1)$.$ The differential forms of the first kind, discussed in the
book by Lang [20], by design do not have any poles while the differentials
of the second kind by design do not have residues. Only differentials of the
third kind have poles of order $\leq $ 1 with nonvanishing residues and,
hence, are physically interesting. We shall be dealing mostly with
differentials of the 2nd kind \ converting them eventually into that of the
third kind. The differentials of the third kind are linearly independent
from that of the first kind according to Lang [20].

Symmetrizing our result, Eq.(3.15), in the spirit of Veneziano ideas we
obtain the 4-particle Veneziano-like amplitude 
\begin{equation}
A(s,t,u)=\tilde{V}(s,t)+\tilde{V}(s,u)+\tilde{V}(t,u)  \tag{3.16}
\end{equation}
where, for example, upon analytical continuation $V(s,t)$ is given by 
\begin{equation}
\tilde{V}(s,t)=(1-\exp (i\frac{\pi }{N}(-\alpha (s))(1-\exp (i\frac{\pi }{N}%
(-\alpha (t))B(\frac{-\alpha (s)}{N},\frac{-\alpha (t)}{N}),  \tag{3.17}
\end{equation}
provided that we have identified $<c_{i}>$ with $\alpha (i),$ $etc.$\ \
Naturally, in arriving at Eq.(3.17) we have extended the differential forms
from that of the second kind to that of the third. The analytical properties
of such designed Veneziano-like amplitudes are discussed in detail below in
subsections on multiparticle amplitudes

\subsection{Connection with CFT trough hypergeometric functions and the
Kac--Moody-Bloch-Bragg condition}

\bigskip

In the light of just obtained results we would like now to compare the
hypergeometric function, Eq.(2.6), with the beta function. Taking into
account that [12] 
\begin{equation*}
(1-zx)^{-a}=\sum\limits_{n=0}^{\infty }\frac{(a,n)}{n!}\left( zx\right) ^{n},
\end{equation*}%
Eq.(2.6) can be rewritten as follows: 
\begin{eqnarray}
F[a,b;c;x] &\dot{=}&\sum\limits_{n=0}^{\infty }\frac{(a,n)}{n!}%
x^{n}\int\limits_{0}^{1}z^{b+n-1}(1-z)^{c-b-1}dz  \notag \\
= &&\sum\limits_{n=0}^{\infty }\frac{(a,n)}{n!}x^{n}B(b+n,c-b). 
\TCItag{3.18}
\end{eqnarray}%
This result is to be compared with Eq.(3.14).To this purpose it is
convenient to rewrite Eq.(3.14) in the following more general form (up to a
constant factor): 
\begin{equation*}
I(m,l)\dot{=}\int\limits_{0}^{1}\unit{u}^{\frac{<c_{1}>-N+mN}{N}}(1-\text{u}%
)^{\frac{<c_{2}>-N+lN}{N}}d\text{u}=B(a+m,b+l),
\end{equation*}%
where $m,l=0,\pm 1,\pm 2,..$ \ It is clear, that the phase factors entering
into Eq.(3.15) will either remain unchanged or will change sign upon such
replacements. At the same time the Veneziano condition, Eq.(3.11a), will
change into 
\begin{equation}
<c_{0}>+<c_{1}>+<c_{2}>=N+mN+lN+kN.  \tag{3.19}
\end{equation}%
This result can be explained physically with help of some known facts from
solid state physics, e.g. read Ref.[29]. To this purpose let us consider the
result of torus action on the form $\omega $, Eq.(3.9). If we demand this
action to be torus action invariant \ (as it is explained in Part 2), then
we obtain, 
\begin{equation}
\sum\limits_{i}<c_{i}>m_{i}=0\func{mod}N,  \tag{3.20}
\end{equation}%
with $m_{i}$ \ being some integers. In particular, consider Eq.(3.20) for a
special case of 4-particle Veneziano amplitude. Then, in \ accord with the
discussion following Eq.(1.5), the Veneziano condition can be rewritten as 
\begin{equation}
<c_{0}>m_{0}+<c_{1}>m_{1}+<c_{2}>m_{2}=0\func{mod}N  \tag{3.21}
\end{equation}%
But, in view of the Griffiths Corollary 2.11, the condition $\func{mod}N$ \
(or \ $\func{mod}2N)$\ \ for the Veneziano amplitudes should actually be
replaced by $N$ (or $2N$). At the same time for the hypergeometric functions
in view of Eq.s(3.19), (3.21), we should write instead

\begin{equation}
<c_{0}>m_{0}+<c_{1}>m_{1}+<c_{2}>m_{2}=mN+lN+kN.  \tag{3.22}
\end{equation}
Such a condition is known in solid state physics as the Bragg equation [29].
This equation plays the central role in determining crystal structure by
X-ray diffraction. Lattice periodicity reflected in this equation affects
kinematics of scattering processes for phonons and electrons in crystals.
Under these circumstances the concepts of particle energy and momentum loose
their usual meaning and should be amended to account for the lattice
periodicity. The same type of amendments should be made when comparing
elementary scattering processes in CFT against those in high energy physics.
We shall call Eq (3.22) the \textit{Kac-Moody-Bloch-Bragg (K--M-B-B) equation%
}. In \ the group-theoretic language of Part 2 the difference between the
high energy \ scattering processes and those in CFT is of the same nature as
the difference between the Coxeter-Weyl (pseudo) reflection groups and their
affine generalizations [30]. The same difference will be explained from the
point of view of \ symplectic and complex manifolds in Part 3.

\subsection{Analytical properties of the multiparticle Veneziano and
Veneziano-like amplitudes (general considerations)}

\bigskip

By analogy with the 4-particle case, the Fermat variety $\mathcal{F}_{aff}$ $%
(N)$ in the affine form in the multiparticle case is given by the following
equation 
\begin{equation}
\text{ }\mathcal{F}_{aff}(N):\text{\ \ }Y_{1}^{N}+\cdot \cdot \cdot
+Y_{n+1}^{N}=1,\text{ }Y_{i}=x_{i}/x_{0}\equiv z_{i}.  \tag{3.23}
\end{equation}
As before, use of parametrization $f:$ $z_{i}=t_{i}^{\frac{1}{N}}\exp (\pm 
\frac{\pi i}{N})$ such that $\sum\nolimits_{i}t_{i}=1$ allows us to reduce
the Fermat variety $\mathcal{F}_{aff}(N)$ to its deformation retract which
is $n+1$ simplex $\Delta .$ I.e. $f($ $\mathcal{F}_{aff}(N))=\Delta $, where 
$\Delta :\sum\nolimits_{i}t_{i}=1$. The period integrals of the type given
by Eq.(3.3) with $\omega $ form given by Eq.(3.9) after taking the
Leray-type residue are reduced to the following standard form (up to a
constant): 
\begin{equation}
I\dot{=}\int\limits_{\Delta }t_{1}^{\frac{<c_{1}>}{N}-1}\cdot \cdot \cdot
t_{n+1}^{\frac{<c_{n+1}>}{N}-1}dt_{1}\wedge \cdot \cdot \cdot \wedge dt_{n},
\tag{3.24}
\end{equation}
where, again, all phase factors have been suppressed temporarily. The
important group-theoretic meaning of the integrand in the above integral
leading to recovery of the model associated with such integral will be
discussed at lenght in both Part 2 and 3 of this work.

For $n=1$ this integral coincides with that given by Eq.(3.14) (up to a
constant) as required. As part of preparations for calculation of this
integral for $n>1$ let us first have another look at the case $n=1$ where we
have integrals of the type 
\begin{equation*}
I=\int\limits_{0}^{1}dxx^{a-1}(1-x)^{b-1}=B(a,b)=\frac{\Gamma (a)\Gamma (b)}{%
\Gamma (a+b)}.
\end{equation*}
Alternatively, we can look at 
\begin{equation}
\Gamma (a+b)I=\int\limits_{0}^{\infty }\int\limits_{0}^{\infty
}dx_{1}dx_{2}x_{1}^{a-1}x_{2}^{b-1}\exp (-x_{1}-x_{2}).  \tag{3.25}
\end{equation}
In the double integral on the r.h.s. let us consider change of variables: $%
x_{1}=\hat{x}_{1}t$ , $x_{2}=\hat{x}_{2}t$ so that $x_{1}+x_{2}=t$ provided
that $\hat{x}_{1}+\hat{x}_{2}=1.$ Taking $t$ and $\hat{x}_{1}$ as new
variables and taking into account that the Jacobian of such transformation
is one the following result is obtained: 
\begin{equation*}
\Gamma (a+b)I=\int\limits_{0}^{\infty }dtt^{a+b-1}\exp
(-t)\int\limits_{0}^{1}d\hat{x}_{1}\hat{x}_{1}^{a-1}(1-\hat{x}_{1})^{b-1},
\end{equation*}
as expected. Going back to the original integral, Eq.(3.24), and introducing
notations $a_{i}=\frac{<c_{i}>}{N}$ we obtain, 
\begin{equation}
\Gamma (\sum\nolimits_{n=1}^{n+1}a_{i})I\dot{=}\int\limits_{0}^{\infty }%
\frac{dt}{t}t^{\sum\nolimits_{i=1}^{n+1}a_{i}}\exp (-t)\int\limits_{\Delta
}t_{1}^{a_{1}-1}\cdot \cdot \cdot t_{n+1}^{a_{n+1}-1}dt_{1}\wedge \cdot
\cdot \cdot \wedge dt_{n}.  \tag{3.26}
\end{equation}
By analogy with the case $n=1$ we introduce new variables :$s_{i}=tt_{i}$.
Naturally, we expect $\sum\nolimits_{n=1}^{n+1}s_{i}=t$ since $t_{i}$
variables are subject to the simplex constraint $\sum%
\nolimits_{i=1}^{n+1}t_{i}=1$.With such replacements we obtain 
\begin{eqnarray*}
\Gamma (\sum\nolimits_{n=1}^{n+1}a_{i})I &=&\int\limits_{0}^{\infty }\cdot
\cdot \cdot \int\limits_{0}^{\infty }\exp
(-\sum\nolimits_{n=1}^{n+1}s_{i})s_{1}^{a_{1}}\cdot \cdot \cdot
s_{n+1}^{a_{n+1}}\frac{ds_{1}}{s_{1}}\wedge \cdot \cdot \cdot \wedge \frac{%
ds_{n+1}}{s_{n+1}} \\
&=&\Gamma (a_{1})\cdot \cdot \cdot \Gamma (a_{n+1}).
\end{eqnarray*}
Using this result, the $n-$particle contribution to the Veneziano amplitude
is given finally by the following expression: 
\begin{equation}
I\dot{=}\frac{\prod\limits_{i=1}^{n+1}\Gamma (a_{i})}{\Gamma
(\sum\nolimits_{n=1}^{n+1}a_{i})}.  \tag{3.27}
\end{equation}

\textbf{Remark 3.1.}\ Eq.(3.27) can be found in paper by Gross [20], page
206, where it is suggested (postulated) without derivation. Eq.(3.27)
provides a complete explicit calculation of the Dirichlet integral,
Eq.(2.8), and, as such, can be found, for example, in the book by Edwards
[31] published in 1922. Calculations similar to ours also can be found in
lecture notes by Deligne [22]. We shall use some additional results from his
notes below.

Our calculations are far from being complete however.To complete our
calculations we need to introduce the appropriate phase factors. In
addition, we need to discuss carefully the analytic continuation \ of just
obtained expression for amplitude to negative values of parameters $a_{i}$.
Fortunately, the phase factors can be reinstalled in complete analogy with
the 4-particle case in view of the following straightforwardly verifiable
identity 
\begin{equation}
B(x,y)B(x+y,z)B(x+y+z,u)\cdot \cdot \cdot B(x+y+...+t,l)=\frac{\Gamma
(x)\Gamma (y)\cdot \cdot \cdot \Gamma (l)}{\Gamma (x+y+...+l)}.  \tag{3.28}
\end{equation}
Because of this identity, the multiphase problem is reduced to that we have
considered already for the 4-particle case and, hence, can be considered as
solved. The analytic continuation problem connected with the multiphase
problem is much more delicate and requires longer explanations.

The first difficulty we encounter is related to the constraints imposed on $%
<c_{i}>$ factors discussed in connection with the 4-particle case, e.g.
restriction :$1\leq <c_{i}>\leq N-1$ $($or $1\leq <c_{i}>\leq N).$ To
resolve this difficulty, we shall follow Deligne's lecture notes [22].We
begin with Eq.(3.9). The Veneziano condition, Eq.(3.11a), extended to the
multivariable case is written as 
\begin{equation}
1=<c>=\frac{1}{N}\sum\limits_{i}<c_{i}>  \tag{3.29}
\end{equation}
whereas the Corollary 2.11. by Griffiths \textit{does not} require this
constraint to be imposed. To satisfy this Corollary, it is sufficient for us
to require only $m=<c>$ for some integer $m$ to be specified below. Clearly,
such a requirement will change the total sum of exponents accordingly in \
the numerator of Eq.(3.9). In particular, for $m=2$ we would reobtain back
Eq.(3.11b). It should be noted at this point that the Lemma 1 by Gross [20]
although imposes such a constraint but was actually proven \textit{not} in
connection with the period differential form, Eq.(3.9). This lemma
implicitly assumes that the Leray residue \textit{was taken already} and
deals with the differential forms occurring as result of such operation. To
avoid guessing in the present case, we need to initiate our analysis again
starting \ from Eq.(3.9) \ and taking into account the Corollary 2.11.

Following Deligne [22], let us discuss what happens if we replace $<c>,$
Eq.(3.7), by $<-c>$. In view of definition of the bracket sign $<...>$ we
obtain, 
\begin{equation}
<-c>=\frac{1}{N}\sum\limits_{i}<-c_{i}>=\frac{1}{N}\sum%
\limits_{i}<-c_{i}+N>=n+2-<c>,  \tag{3.30}
\end{equation}
where the factor $n+2$ comes from the sum $\sum\nolimits_{i}1$ and $<c>$ is
the same as in Eq.(3.7), provided that $1\leq <c_{i}>\leq N.$ This result
implies that the number $m$ defined above can be only in the range 
\begin{equation}
\frac{n+2}{N}\leq m\leq n+2.  \tag{3.31}
\end{equation}

\textbf{Remark 3.2}. The Fermat variety $\mathcal{F}(N)$, Eq.(3.6), is of
the \textit{Calabi-Yau} type if and only if $n+2=N$ , Ref.[32], page 531.
Clearly, this requirement is equivalent to the Veneziano condition,
Eq.(3.29), i.e. $m=1$.

\textbf{Remark} \textbf{3.3}. By \textit{not} imposing this condition we can
get still many physically relevant interesting results using Deligne's
lecture notes [22].We have encountered this already while arriving at
Eq.(3.11b). Clearly, this equation is anyway reducible to Eq.(3.11a) but
earlier we obtained physically important phase factor $\varepsilon $
(instead of $\xi )$ by working with Eq.(3.11b). It should be obvious by now
that $m$ is responsible for change in phase factors: from $\xi $ ( for $m=1$%
)-to $\varepsilon $ $($ for $m=2$)-to $\hat{\varepsilon}_{m}=\exp (i\frac{%
2\pi }{mN})$ ( for $m>2$).$\ $Physical significance of these phase factors
is discussed in the next subsection.

To extend these results we need to introduce several new notations now. Let $%
V_{\mathbf{C}}$ be finite dimensional vector space over \textbf{C}. A 
\textbf{C}-rational Hodge structure of weight $n$ on $V$ is a decomposition $%
V_{\mathbf{C}}=\bigoplus\limits_{p+q=n}V^{p,q}$ such that $\bar{V}%
^{p,q}=V^{q,p}$ \ We extend the definition of the torus action \ (to be
given\ rigorously in Part 2) in order to accommodate the complex conjugation
: $(t,V^{p,q})$=$t^{-p}\bar{t}^{-q}V^{p,q}.$ Next, we define the filtration
(the analog of the flag decomposition, e.g. see Ref. [33] or our earlier
work, Ref.[15], and/or Part 2 for rigorous definitions and further
discussion) via $F^{p}V=\bigoplus\limits_{p^{\prime }>p}V^{p^{\prime
},q^{\prime }}$ so that $\cdot \cdot \cdot \supset F^{p}V\supset
F^{p+1}V\supset \cdot \cdot \cdot $ is a decreasing filtration on $V$. The
differential form, Eq.(3.9), belongs to the space $\Omega _{m}^{n+1}(%
\mathcal{F})$ of differential forms such that $\omega =\dfrac{p(\mathbf{z})}{%
q(\mathbf{z})^{m}}\omega _{0}$ , where $p(\mathbf{z})$ is a homogenous
polynomial of degree $m\deg (q)-(n+2).$ Such differential forms have a pole
of order $\leq m.$ As in the standard \ complex analysis, one can define the
multidimensional analogue of the residue via map $R(\omega )$: $\Omega
_{m}^{n+1}(\mathcal{F})\rightarrow $H$^{n}(\mathcal{F},\mathbf{C})$ via 
\begin{equation}
<\sigma ,R(\omega )>=\frac{1}{2\pi i}\int\nolimits_{\sigma }\omega \text{ , }%
\sigma \in \text{H}_{n}(\mathcal{F},\mathbf{C}).  \tag{3.32}
\end{equation}
Deligne proves that:

a) H$^{n}(\mathcal{F},\mathbf{C})=\bigoplus\limits_{\bar{c}\neq 0}$H$^{n}(%
\mathcal{F},\mathbf{C})_{\bar{c}}$ , where

b)H$^{n}(\mathcal{F},\mathbf{C})_{\bar{c}}\subset F^{<c>-1}$H$^{n}(\mathcal{F%
},\mathbf{C}),$ \ while the complex conjugate of H$^{n}(\mathcal{F},\mathbf{C%
})_{\bar{c}}$

\ \ \ is given by H$^{n}(\mathcal{F},\mathbf{C})_{-\bar{c}}\subset
F^{n-<c>+1}$H$^{n}(\mathcal{F},\mathbf{C}).$

Thus, by construction, H$^{n}(\mathcal{F},\mathbf{C})_{\bar{c}}$ is of
bidegree \ $(p,q)$ with $p=<c>-1,$ $q=n-p,$ while its complex conjugate H$%
^{n}(\mathcal{F},\mathbf{C})_{-\bar{c}}$ is of bidegree $(q,p).$ Obtained
cohomologies are nontrivial and of Hodge-type only when $<c>\neq 1.$
Finally, the procedure of extracting the residue from integral in Eq.(3.32)
with $\omega $ containing a pole of order $m$ is described in the book by
Hwa and Teplitz [28] and in spirit is essentially the same as in the
standard one -variable complex analysis. Therefore, after all, we end up
again with the differential form $\omega $, Eq.(3.9), with $<c>=1.$ However,
this form will be used with the phase factors $\hat{\varepsilon}_{m}$
instead of $\xi .$ Physical consequences of this replacement are considered
in the next subsection.

\bigskip

\subsection{Analytical properties of the Veneziano-like amplitudes
(ramifications)}

\bigskip

Earlier obtained results of \ this section allow us to write the following
4- particle Veneziano-like amplitude, 
\begin{equation}
A(s,t,u)=\tilde{V}(s,t)+\tilde{V}(s,u)+\tilde{V}(t,u),  \tag{3.16}
\end{equation}
where, for instance, 
\begin{equation}
\tilde{V}(s,t)=(1-\exp (i\frac{\pi }{N}(-\alpha (s))(1-\exp (i\frac{\pi }{N}%
(-\alpha (t))B(-\frac{\alpha (s)}{N},-\frac{\alpha (t)}{N}).  \tag{3.33}
\end{equation}
Although this result was obtained by the same analytic continuation as in
the case of the Veneziano amplitude, the \ resulting analytical properties
of such Veneziano-like amplitude are markedly different. In this subsection
we would like to discuss these important differences.

We begin by noticing that, in view of Eq.(3.11b), the Veneziano condition in
its simplest form: $a+b+c=1,$ upon analytic continuation, leads again to the
requirement: $\alpha (s)+\alpha (t)+\alpha (u)=-1,$ if we identify, for
example, $\frac{<c_{1}>}{N}$ $=a_{1}$\ with -$\alpha (s),$etc. This naive
identification leads to some difficulties however. Indeed, since physically
we are interested in the poles and zeros of gamma functions, we expect our
parameters $a,b$ and $c$ to be integers. This is possible only if the
absolute values of $\ <c_{1}>,<c_{2}>$ and $<c_{3}>$ are greater or equal
than $N$. By allowing these parameters to become greater than $N$ we would
formally violate the requirements of Corollary 2.11. by Griffiths, e.g. see
Eq.(3.9) and the discussion around it. Fortunately, the occurring difficulty
can by resolved. For instance, one can postulate Eq.s (3.16),(3.33) as 
\textit{defining relations } for the Veneziano-like amplitudes as it was
done historically by Veneziano for what has become known as the Veneziano
amplitudes. In this case one is confronted with the problem of finding some
physical model reproducing such amplitudes. To facilitate search for such a
model it is reasonable to impose the same constraints as for the standard
Veneziano amplitudes. Clearly, if we want to use earlier obtained results,
we must, in addition to these constraints, to impose the the constraint
coming from the Corollary 2.11. This Corollary formally forbids us from
consideration of ratios $<c_{i}>/N$ \ whose absolute value is greater than
one as we have discussed in the previous subsection. \ After a short while
of thinking, this complication creates no additional problems however. This
can be seen already on example of Eq.(1.15) for $\sin \pi x$. Indeed,
consider the function 
\begin{equation*}
F(x)=\frac{1}{\sin \pi x}.
\end{equation*}
It will have the first order poles for $x=0,\pm 1,\pm 2,...$ If we formally
define the bracket operator $<...>$ by analogy with that defined ibefoe
Eq.(3.7), e.g. $\ 0<<x>\leq 1$ $\forall x,$ then to reproduce the poles of $%
F(x)$ it is sufficient to write 
\begin{equation}
F(x)=\frac{1}{\sin \pi x}=\frac{1}{1-<x>}.  \tag{3.34}
\end{equation}
Clearly, the above result can be read as well from right to left, i.e.
removal of brackets, is equivalent to unwrapping $S^{1}$ into $R^{1}$, i.e.
to swithching from a given space, e.g. $S^{1},$ to its universal covering
space, e.g. $\mathbf{R}^{1}$. \ By looking at Eq.(1.16) for expression for $%
\Gamma (z)$ and comparing it with Eq.(1.15) we notice that all singularities
of $\Gamma (z)$ are exactly the same as those for $F(x)$. Hence, the same
unwrapping is applicable for this case as well. These observations lead us
to the following set of prescriptions:

a) use Eq.(3.15) in Eq.(3.16) in order to obtain the full Veneziano-like

\ \ amplitude,

b) remove brackets,

c) perform analytic continuation to negative values of $c_{i}^{\prime }s$ ,

d) identify $-c_{i}/N$ with $-\alpha (i),($ $i=s$, $t$ or $u$).

After this, let, for instance, $\alpha (s)=a+bs$ where both $a$ and $b$ are
some positive (or better, non negative) constants. Then, the tachyonic pole: 
$\alpha (s)=0,n=0$ \ (e.g. see Eq.s(1.17)-(1.19) ) is killed by the
corresponding phase factor in Eq.(3.33). The mass spectrum is determined by:
a) the actual numerical values of the constants $a$ and $b,$ b) by the phase
factors and c) by the values of parameter $N$ (even or odd).

For instance, the condition, Eq.(3.11b), leads to the requirement that the
particle with masses satisfying equation $\alpha (s)=2l,l=1,2,...$ cannot be
observed since the emerging pole singularities are killed by \ zeroes coming
from the phase factor. In the case of 4-particle amplitude the inequality,
Eq.(3.31), should be used with $n=1$ thus leading to the constraints: $N\geq
3$ and $1\leq m\leq 3.$ If we choose $m=3$ we obtain similar requirement
forbidding particles with masses coming from the equation $\alpha
(s)=3l,l=1,2,...$.\footnote{%
E.g. see Remark 3.3.}

Such limitations are not too severe, however. Indeed, let us consider for a
moment the existing bosonic string parameters associated with the Veneziano
amplitude. For the open string the known convention is: $\alpha (s)=1+\frac{1%
}{2}s.$ The tachyon state is determined therefore by the condition: $\alpha
(s)=0$ thus producing $s=-2$ $=M^{2}$. \ If now $1+\frac{1}{2}s=l,$ we
obtain: $s=2(l-1),$ $l=1,3,5,..$.(for $m=2$) or $l=1,2,4,5,..$.(for $m=3$).
Clearly, the combined use of these results produce the mass spectrum for the
open bosonic string (without tachyons). If we want the graviton to be
present in the spectrum we have to adjust the values of constants $a$ and $b$%
. For instance, it is known [2] that for the closed string the tachyon
occurs at $s=-8$ $=M^{2}$. This result can be obtained if we choose either $%
\alpha (s)=2+\frac{1}{4}s$ or $\alpha (s)=1+\frac{1}{8}s.$ To decide which
of these two expressions provides better fit to the experimental data we
recall that the massless graviton should have spin equal to two. If we want
the graviton to be present in the spectrum we must select the first option.
This is so because of the following arguments. First, we have to take into
account that for large $s$ and fixed $t$ the amplitude $V(s,t)$ can be
approximated by (Ref.[2], page 10), 
\begin{equation}
V(s,t)\simeq \Gamma (-\alpha (t))(-\alpha (s))^{\alpha (t)}  \tag{3.35}
\end{equation}%
while the Regge theory predicts (Ref. [2], pages 3,4) that 
\begin{equation}
V_{J}(s,t)=-\frac{g^{2}(-s)^{J}}{t-M_{J}^{2}}\simeq \frac{-g^{2}(-\alpha
(s))^{J}}{\alpha (t)-J}  \tag{3.36}
\end{equation}%
for the particle with spin $J$. This leaves us with the first option.
Second, by selecting this option our task is not complete since so far we
have ignored the actual value of the Fermat parameter $N$. Such ignorance
causes emergence of the fictitious tachyon coming from the equation $2+\frac{%
1}{4}s=l$ for $l=1$.This difficulty is easily resolvable if we take into
account that the \textquotedblright Shapiro-Virasoro\textquotedblright\
condition, Eq.(3.11b), is reducible to the \textquotedblright Veneziano
condition, Eq.(3.11a), in the case if all $<c_{i}>$ in Eq.(3.11b) are even.
At the same time, if $N$ in Eq.(3.11a) is even, it can be brought to the
form of Eq.(3.11b). Hence, in making identification of $-c_{i}/N$ with $%
-\alpha (i)$ we have to consider two options:

a) $N$ is odd, then $c_{i}/N=\alpha (i)$, b) $N$ is even, $N=2\hat{N}$ ,
then $c_{i}/\hat{N}=\alpha (i).\footnote{%
Then, proceeding by analogy with arguments for the open bosonic string
spectrum, we reobtain the spectrum of closed bosonic string}$

The fictitious tachyon is removed from the spectrum in the case if we choose
the option b). Clearly, after this, in complete analogy with the ''open
string'' case, we re obtain the tachyon-free spectrum of the ''closed''
bosonic string.

To complete our investigation of the Veneziano-like amplitudes we still
would like to have some discussion related to Eq.s(3.35),(3.36). To this
purpose, using integral representation of $\Gamma $ given by 
\begin{equation*}
\Gamma (x)=\int\limits_{0}^{\infty }\frac{dt}{t}t^{x}\exp (-t)
\end{equation*}
and assuming that $x$ is large and positive, the leading term of the saddle
point approximation (to $\Gamma )$ is readily obtained, and is given by 
\begin{equation*}
\Gamma (x)=Ax^{x}\exp \left( -x\right) ,
\end{equation*}
where $A$ is some constant. Applying \ (with some caution) this result to 
\begin{equation*}
V(s,t)=\frac{\Gamma (-\alpha (s))\Gamma (-\alpha (t))}{\Gamma (-\alpha
(s)-\alpha (t))}\equiv B(-\alpha (s),-\alpha (t))
\end{equation*}
we obtain Eq.(3.35).

Although such arguments formally explain the origin of the Regge asymptotic
law, Eq.(3.35), they do not illuminate the combinatorial origin of this
result essential for its generalization. To correct this deficiency we would
like to use\ again Eq.(1.21). In the case when $z_{0}=...=z_{n}=1$ the inner
sum in the right hand side yields the total number of monomials of the type $%
z_{0}^{k_{0}}\cdot \cdot \cdot z_{n}^{k_{n}}$ with $k_{0}+...+k_{k}=n$. The
total number of such monomials is given by Eq.(1.22) which allows us to
write the generating function $P(k,t)$, Eq.(1.23), and, accordingly, the
Veneziano amplitude, Eq.(1.24). The function $p(k,n)$ defined in Eq.s (1.22)
and (1.26) happen to be the non negative integer. In Section 1 we have
mentioned already that it arises naturally as the dimension of the quantum
Hilbert space associated (through the coadjoint orbit method) with the
symplectic manifold of dimension $2k$ constructed by ''inflating'' $\Delta
_{k}.$ \ We would like to use these observations now to complete our
discussion of the Regge-like result, Eq.(3.35). To this purpose using
Eq.(1.14) and assuming that $\alpha (t)\rightarrow k^{\ast }$ we can
approximate the amplitude $V(s,t)$ by 
\begin{equation}
V(s,t)\simeq -\frac{p_{\alpha (s)}(k^{\ast })}{\alpha (t)-k^{\ast }}\simeq -%
\frac{p_{\alpha (s)}(\alpha (t))}{\alpha (t)-k^{\ast }}.  \tag{3.37}
\end{equation}
For large $\alpha (s)\footnote{%
Notice that the negative sign in front of $\alpha (s)$ was already taken
into account}$ by combining Eq.s(1.26) and (3.37) we obtain, 
\begin{equation}
V(s,t)\simeq \frac{-1}{\alpha (t)-k^{\ast }}\frac{\alpha (s)^{k^{\ast }}}{%
k^{\ast }!}.  \tag{3.38}
\end{equation}
In view of the footnote remark, and taking into account that $k^{\ast
}\simeq \alpha (t),$ this result coincides with Eq.(3.35) as required. In
addition, however, for the large $k^{\prime }s$ it can be further rewritten
as

\begin{equation}
V(s,t)\simeq \frac{-1}{\alpha (t)-k^{\ast }}(\frac{\alpha (s)}{k^{\ast }}%
)^{k^{\ast }}\simeq \frac{-1}{\alpha (t)-k^{\ast }}(\frac{\alpha (s)}{\alpha
(t)})^{\alpha (t)}  \tag{3.39}
\end{equation}
in accord with Eq.(3.36). Obtained result is manifestly symmetric with
respect to exchange $s\rightleftharpoons t$ in accord with earlier mentioned
requirement $V(s,t)=V(t,s)$. Moreover, it explicitly demonstrates that the
angular momentum of the graviton is indeed equal to two as required.

\bigskip

\pagebreak

\bigskip

\bigskip

\textbf{References}

\bigskip

\bigskip

[1] \ \ G.Veneziano, Construction of crossing symmetric, Regge

\ \ \ \ \ \ \ behaved, amplitude for linearly rising trajectories, Il Nuovo

\ \ \ \ \ \ \ Chim. 57A (1968) 190-197

\bigskip

[2] \ \ \ M.Green, J.Schwarz, E.Witten, Superstring Theory. Vol.1.,

\ \ \ \ \ \ \ Cambridge University Press, Cambridge (UK), 1987

\bigskip

[3] \ \ \ R.Stanley, Combinatorics and Commutative Algebra\textit{.},

\ \ \ \ \ \ \ \ Birkh\"{a}user, Inc., Boston, 1996

\bigskip

[4] \ \ \ M.Virasoro, Alternative construction of crossing-symmetric

\ \ \ \ \ \ \ \ amplitudes with Regge behavior, Phys. Rev. 177 (1969)
2309-2314

\bigskip

[5] \ \ \ \ A.Kholodenko, New string amplitudes from old Fermat
(hyper)surfaces,

\ \ \ \ \ \ \ IJMP A \ 19 (2004) 1655-1703

\bigskip

[6] \ \ \ V.De Alfaro, S. Fubini, G. Furlan, C. Rossetti, Currents in Hadron

\ \ \ \ \ \ \ \ Physics, Elsevier Publ.Co., Amsterdam, 1973

\ 

[7] \ \ \ F.Hirzebruch, D. Zagier, The Atiyah-Singer Theorem and

\ \ \ \ \ \ \ Elemetary Number Theory, Publish or Perish Inc.,

\ \ \ \ \ \ \ Berkeley, Ca, 1974

\bigskip

[8] \ \ \ \ R. Stanley, Enumerative Combinatorics,Vol.1,

\ \ \ \ \ \ \ \ Cambridge University Press, Cambridge, UK, 1997

\bigskip

[9] \ \ \ \ M. Vergne, Convex polytopes and quanization of symplectic
manifolds,

\ \ \ \ \ \ \ \ Proc.Natl.Acad.Sci. 93 (1996) 14238-14242

\bigskip

[10] \ \ \ P. Etingof, I. Frenkel, A. Kirillov. Jr., Lectures on
Representation

\textit{\ \ \ \ \ \ \ \ }Theory and Knizhnik-Zamolodchikov Equations\textit{,%
}

\ \ \ \ \ \ \ \ AMS, Providence, RI, 1998

[11] \ \ \ I.Gelfand, M. Kapranov, A. Zelevinsky, Generalized Euler

\ \ \ \ \ \ \ \ \ integrals and A-hypergeometric functions, Adv. in Math. 84
(1992)

\ \ \ \ \ \ \ \ \ 255-271 ; ibid \ 96 (1992) 226-263

[12] \ \ \ P. Orlik, H. Terrao, Arrangements and Hypergeometric Integrals%
\textit{,}

\ \ \ \ \ \ \ \ \ Math.Soc.Japan Memoirs, Vol.9, Japan Publ.Trading Co.,

\ \ \ \ \ \ \ \ \ Tokyo, 2001

\bigskip

[13] \ \ \ \ J. Milnor, On the 3-dimensional Brieskorn manifolds M(p,q,r),

\ \ \ \ \ \ \ \ \ \ In: Knots, Groups, and 3-Manifolds (Papers dedicated to

\ \ \ \ \ \ \ \ \ \ the memory of R. H. Fox), pp. 175--225,

\ \ \ \ \ \ \ \ \ \ Ann. of Math. Studies. 84.

\ \ \ \ \ \ \ \ \ \ Princeton Univ. Press, Princeton, 1975

\bigskip

[14] \ \ \ P.Deligne, G.Mostow, Commensurabilities Among Lattices

\ \ \ \ \textit{\ \ \ \ \ }in PU (1, n), Ann. of Math. Studies. 132,

\ \ \ \ \ \ \ \ \ Princeton Univ. Press, Princeton, 1993

\bigskip

\ [15] \ \ \ \ A. Kholodenko, Kontsevich-Witten model from 2+1 gravity:

\ \ \ \ \ \ \ \ \ \ \ new exact combinatorial solution,

\ \ \ \ \ \ \ \ \ \ \ J.Geom.Phys. 43 (2002) 45-91 (2002)

\bigskip

\ [16] \ \ \ R. Stanley, Enumerative Combinatorics,Vol.2,

\ \ \ \ \ \ \ \ \ Cambridge University Press, Cambridge (UK), 1999

\bigskip

[17] \ \ \ S. Chowla, A.Selberg, On Epstein's Zeta -function,

\ \ \ \ \ \ \ \ \ J.Fur die Reine und Angev.Math. 227 (1967) 86-100

\bigskip

[18] \ \ \ A.Weil, Sur les periods des integrales Abeliennes,

\ \ \ \ \ \ \ \ \ Comm. Pure and appl.Math. 29 (1976) 813-819

\bigskip

[19] \ \ \ A.Weil, Abelian varieties and the Hodge ring,

\ \ \ \ \ \ \ \ \ In: Collected Works, Vol.3, Springer-Verlag, Berlin, 1979

\bigskip

[20] \ \ \ B.Gross, On periods of Abelian integrals and formula of

\ \ \ \ \ \ \ \ \ Chowla and Selberg, Inv.Math. 45 (1978) 193-211

\bigskip

[21] \ \ \ S. Lang, Introduction to Algebraic and Abelian Functions.

\ \ \ \ \ \ \ \ \ Springer-Verag, Inc., Berlin, 1982

\bigskip

[22] \ \ \ P.Deligne, Hodge cycles and Abelian varieties,

\ \ \ \ \ \ \ \ \ In: \ Lecture notes in Math. 900,

\ \ \ \ \ \ \ \ \ Springer-Verlag, Berlin, 1982

\bigskip

[23] \ \ \ J.Carson, S. Muller-Stach, C.Peters, Period Mappings and Period

\ \ \textit{\ \ \ \ \ \ \ }Domains, Cambridge University Press, Cambridge
(UK) 2003

\bigskip

[24] \ \ \ P. Griffiths, On periods of certain rational integrals,

\ \ \ \ \ \ \ \ \ Ann.of Math. 90 (1969) 460-495; ibid 495-541

\bigskip

[25] \ \ \ Y.Manin, Algebraic curves over fields with differentiation.

\ \ \ \ \ \ \ \ \ AMS Translations 206 (1964) 50-78

\bigskip

[26] \ \ \ D.Rohrlich, Points at infinity on the Frermat Curves,

\ \ \ \ \ \ \ \ \ Inv.Math. 39 (1977) 95-127

\bigskip

[27] \ \ \ J. Leray, Le calcul differential et integral sur une variete

\ \ \ \ \ \ \ \ \ analytique complexe, Bull.Soc.Math.France 57 ( 1959) 81-180

\bigskip

[28] \ \ \ R.Hwa, V.Teplitz, Homology and Feynman Integrals,

\ \ \ \ \ \ \ \ \ W.A.Benjamin, Inc., New York, 1966

\bigskip

[29] \ \ N.Ashcroft, D. Mermin, Solid State Physics,

\ \ \ \ \ \ \ \ Saunders College Press, Philadelphia, 1976

\bigskip

[30] \ \ J.Humphreys, Reflection Groups and Coxeter Groups,

\ \ \ \ \ \ \ \ Cambridge University Press, cambridge (UK), 1990

\bigskip

[31] \ \ J. Edwards, Tretease on the Integral Calculus\textit{, }Vol.2%
\textit{\ ,}

\ \ \ \ \ \ \ \ Macmillan Co., London,1922

\bigskip

[32] \ \ \ N. Yui, Arithmetics of certain Calabi-Yau varieties

\ \ \ \ \ \ \ \ and mirror symmetry.

\ \ \ \ \ \ \ \ In: B.Conrad, K.Rubin (Eds). Arithmetic Algebraic Geometry.

\ \ \ \ \ \ \ \ AMS, Providence (RI), 2001

\bigskip

[33] \ \ \ W.Fulton, Young Tableaux,

\ \ \ \ \ \ \ \ \ Cambridge University Press, Cambridge (UK), 1997

\ \ \ \ \ \ \ \ \ \ \ 

\bigskip

\end{document}